%% file: twist-pr.tex
\documentclass[aps,twocolumn,floatfix,pre]{revtex4-2}

\input{preamble.tex}

\begin{document}

\title{Magic angle twisted bilayer graphene as a highly efficient quantum Otto engine}
\author{Ayush Singh}
\email{ayush.singh@niser.ac.in}
\affiliation{School of Physical Sciences,
National Institute of Science Education and Research, HBNI, Jatni 752050, India}

\author{Colin Benjamin}
\email{colin.nano@gmail.com}
\affiliation{School of Physical Sciences,
National Institute of Science Education and Research, HBNI, Jatni 752050, India}

\begin{abstract}
At a discrete set of \emph{magic angles}, twisted bilayer
graphene has been shown to host extraordinarily flat bands, correlated
insulating states, unconventional superconductivity, and distinct Landau
level degeneracies. In this work, we design a highly efficient quantum Otto engine using a twisted bilayer graphene sample. Flat electronic bands at magic angles make the prospect of extracting work from our Otto engine lucrative because exploiting correlated phenomena may lead to nanoscale devices are operating at more considerable efficiencies.
We use an eight-band continuum model of twisted bilayer
graphene to compute efficiencies and work outputs for magic and non-magic angle twists and compare the results with an $AB$ stacked bilayer and a monolayer. It is observed that the efficiency varies smoothly with the twist angle, and the maximum is attained at the magic angle.
\end{abstract}

\maketitle

\section{Introduction}

One of the ways to approach quantum thermodynamics \cite{binder18} is to design and study thermodynamic cycles designed as quantum analogs of classical thermodynamic processes. These cycles use quantum matter as a ``working substance'' to convert thermal energy to valuable work. In the literature, quantum analogs of adiabatic, isochoric, isothermal~\cite{quan07} and isobaric~\cite{quan09} process have been described, and some general results with quantum thermodynamic cycles, like Carnot and Otto, have been derived~\cite{rezek06, quan07, quan09, quan05, uzdin14, vinjanampathy15}.

Due to the quantum nature of the working substance, quantum heat engines (QHEs) are expected to exhibit some unique properties that allow better performance than their classical counterparts. For example, it has been shown that quantum heat engines can extract work from a single heat bath~\cite{scully01}, and under certain conditions even surpass the Carnot limit~\cite{rossnagel14, scully11}. Hence, QHEs can be used to convert thermal energy to practical work in nanoscale devices efficiently. In addition to this, studying quantum thermodynamic cycles allows us to test the robustness of thermodynamic principles in a quantum setting. Moreover, since the language used to describe QHEs is quite general, the same discussion can be applied to phenomena ranging from lasers and photosynthetic light harvesting~\cite{scully03, scully11, dorfman13, alicki17} to information theory and quantum computation~\cite{kieu04, alicki04, toyabe10}.

Quantum thermodynamic processes are carried out either by quasistatically changing the temperature of the heat reservoir---which the working substance is kept in equilibrium with---or by varying some tunable parameter that changes the energy spectrum of the quantum system. In magnetically driven quantum heat engines, Landau levels are changed by varying an external magnetic field \cite{munoz14, pena19, pena20, pena20b}. It is convenient because it is generally easier to quasistatically modulate the external field than some internal parameter of the working substance \cite{munoz14}. Magnetically driven quantum heat engines based on a semiconductor quantum dot~\cite{munoz14,
pena19} and monolayer graphene flake~\cite{pena15,pena20} have been
proposed, however, these are by no means the only kinds of QHEs
possible.

In this paper, we propose a magnetically driven quantum heat engine based on
twisted bilayer graphene. Interest in twisted bilayer graphene (TBG) and other
Moir\'e materials have exploded in the last few years because of band topology, electronic
and optical properties of Moir\'e systems can be tuned by
engineering the relative twist between layers.
In particular, at a discrete set of magic angles, TBG hosts
exceptionally flat electronic bands where Fermi velocity vanishes and the
two layers get strongly coupled~\cite{bistritzer11a, dossantos12}. Flat bands
are interesting because they can lead to highly correlated phenomena such as
superconductivity~\cite{cao18b}, insulating states at
half-filling~\cite{cao18a}, isospin Pomeranchuk effect~\cite{saito21} and an
electronic phase transition at zero magnetic field~\cite{rozen21}. A quantum
heat engine based on twisted bilayer graphene is, therefore, a great avenue to
study the interplay of electronic properties of Moir\'e systems with quantum
thermodynamics, statistical physics, and quantum information.

In this paper, we present calculations for a quantum analog of the Otto cycle
\cite{quan05, quan07, abah12} based on bilayer graphene, and observe that the
efficiency increases when the layers are twisted with respect
to each other and approaches the maximum at the magic angle. For computing
Landau levels in TBG, we use an eight-band approximation of the non-interacting
continuum model Hamiltonian \cite{dossantos07, bistritzer11a, bistritzer11b,
dossantos12} which reproduces the Fermi velocity with reasonable accuracy down
to the first magic angle~\cite{bistritzer11a}, and diagonalize it
numerically~\cite{python19}.

The rest of this paper is organized as follows: we start by reviewing Landau levels in magic-angle twisted graphene (MATBG), followed by a description of the Otto engine cycle. Expressions for work output and efficiency are derived, and the results are compared for monolayer, bilayer, and magic-angle twisted bilayer.

\section{Landau levels in monolayer and bilayer graphene}

Since our proposed quantum heat engine uses a graphene flake under
transverse magnetic field as working substance, we present here a brief review
of Landau levels in monolayer and bilayer graphene. The treatment here closely
follows \cite{goerbig11} for monolayer and \cite{mccann13} for bilayer
graphene.

Effective low energy Hamiltonian for monolayer graphene near valley points
is
\begin{equation}
h_m(\bm{k}) = \xi \hbar v_F \bm{\sigma}\cdot\bm{k},
\label{e:graphene-hamiltonian}
\end{equation}
where $\xi = \pm$ is the \emph{valley pseudospin}, $v_F \sim
\SI{e6}{\meter\per\second}$ is Fermi velocity, $\bm{\sigma} = (\sigma_x,
\sigma_y)$ are Pauli matrices, and $\bm{k} = (k_x, k_y)$ is crystal
momentum. This Hamiltonian leads to massless Dirac fermions
with Berry phase~$\pi$ \cite{goerbig11}. In order to incorporate magnetic
field we use the gauge transformation $\bm{p} \to \bm{\pi} = \bm{p} + e\bm{A}$,
where $\bm{A}$ is magnetic vector potential and charge of electron is
$-e$. For a transverse magnetic field $\bm{B} = (0, 0, B)$,
vector potential in Landau gauge becomes $\bm{A} = (0, Bx, 0)$ which results in
$\pi_x = p_x$ and $\pi_y = p_y + eBx$.
With canonical commutation relations $[x_i, p_j] = i\hbar \delta_{ij}$, it
can be shown that the operators,
\begin{equation}
\Pi = \frac{1}{\sqrt{2e\hbar B}} \pqty{\pi_x - i\pi_y}
\quad\text{and}\quad
\Pi^\dagger = \frac{1}{\sqrt{2e\hbar B}} \pqty{\pi_x + i\pi_y},
\end{equation}
satisfy the algebra of harmonic oscillator ladder operators i.e., $[\Pi,
\Pi^\dagger] = 1$. In terms of these ladder operators the
Hamiltonian~(\ref{e:graphene-hamiltonian}) can be written as,
\begin{equation}
h_m = \xi \sqrt{2} \frac{\hbar v_F}{l_B} \begin{bmatrix} 0 & \Pi \\
\Pi^\dagger & 0 \end{bmatrix},
\label{e:ladder-graphene-hamiltonian}
\end{equation}
where we have introduced \emph{Landau radius} $l_B = \sqrt{\hbar/eB}$.
Eigenvalue equation for~(\ref{e:ladder-graphene-hamiltonian}) can be solved
exactly to give Landau levels for monolayer graphene~\cite{goerbig11},
\begin{equation}
E_n = \pm \frac{\hbar v_F}{l_B}\sqrt{2n},\quad n = 1, 2, 3, \ldots
\end{equation}
where $\pm$ is the \emph{band index} labelling conduction and valence bands,
and $n$ is Landau level index.
We neglect Zeeman splitting and note that each $n$-state is fourfold
degenerate due to spin and valley degeneracies. For $n = 0$, we get a
fourfold degenerate ground state at zero energy.

In tight-binding model for $AB$ stacked bilayer graphene, there are four
nearest-neighbour tunnelling processes: intralayer hopping, dimer hopping and
two non-dimer hoppings, and we
get massive Dirac fermions with Berry phase $2\pi$ \cite{mccann06, mccann13}.
If we only consider intralayer and dimer hoppings in tight-binding model, a
low energy Hamiltonian can be derived,
\begin{equation}
h_b(\bm{k}) = -\frac{1}{2m_\text{eff}}\mqty[ 0 & \Pi^2 \\
(\Pi^\dagger)^2 & 0],
\label{e:bilayer-ham}
\end{equation}
which admits an analytical solution for
Landau levels,
\begin{equation}
E_n = \pm \hbar \omega_B \sqrt{n (n - 1)},\quad n = 2, 3, 4, \ldots
\label{e:bilayer-ll}
\end{equation}
where $\omega_B = eB/m_{\text{eff}}$ is \emph{cyclotron frequency} with
effective mass $m_{\text{eff}} \approx 0.035 m_e$ \cite{mccann06, mccann13}.
Like in case of monolayer, each $n$ state is fourfold degenerate due to
spin and valley degeneracies, but both $n = 0$ and $n = 1$ are zero energy
states and ground state is therefore eightfold degenerate. This
spectrum is valid only for small level index and low magnetic fields because,
in obtaining (\ref{e:bilayer-ham}), the trigonal warping term due to $\gamma_3$
was dropped, and orbitals relating to dimer sites due to $\gamma_4$, were
eliminated \cite{mccann13}. In particular, we require $n\hbar\omega_B \ll
\SI{3}{eV}$, which is easy to satisfy for a heat engine operating around
$\SI{100}{K}$ in which only first few Landau levels are occupied.

\subsection{Model for twisted bilayer graphene}

\begin{figure}[ht]
\centering
\includegraphics[width=0.3\textwidth]{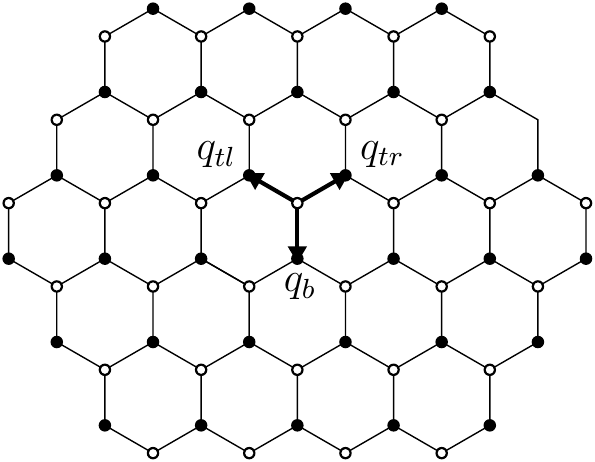}
\caption{\label{fig:lattice} Momentum space lattice of twisted bilayer
graphene. Three equivalent Dirac points result in three distinct
tunnelling processes. For all three processes $\abs{\bm{q}_j} =
k_\theta$ and the hopping directions are: $(0, -1)$ for
$\bm{q}_\text{b}$, $(\sqrt{3}/2, 1/2)$ for $\bm{q}_\text{tr}$ and
$(-\sqrt{3}/2, 1/2)$ for $\bm{q}_\text{tl}$.}
\end{figure}

The low energy continuum model Hamiltonian for twisted bilayer graphene
consists of three parts: two single layer Hamiltonians for intralayer hopping
and a term for tunnelling between layers \cite{dossantos07, bistritzer11a,
dossantos12}. The single layer Hamiltonian, rotated by an angle $\theta$ for an
isolated graphene sheet near Dirac point is
\begin{equation}
h_\theta(\bm{k}) = \mathcal{D}(\hat{\bm{z}}, \theta) \big[-\hbar v_F
\bm{\sigma}\cdot\bm{k}\big]\mathcal{D}\inv(\hat{\bm{z}}, \theta),
\label{e:twisted-graphene-hamiltonian}
\end{equation}
where $\bm{k} = (k_x, k_y)$ is crystal momentum, $\bm{\sigma} = (\sigma_x,
\sigma_y)$ are Pauli matrices and $\mathcal{D}(\hat{\bm{z}}, \theta) =
e^{-i\sigma_z \theta/2}$ is the rotation matrix. Dirac points
of the two rotated graphene layers are separated by $k_\theta =
(8\pi/3a)\sin(\theta/2)$, where $a = \SI{2.46}{\angstrom}$ is the lattice
constant~\cite{bistritzer11a}. For interlayer tunnelling, an analysis of
Moir\'e patterns shows that, for small twist angles there are three main
tunnelling processes, with hopping directions (cf. Fig.~\ref{fig:lattice})
\begin{equation}
\bm{q}_\text{b} = k_\theta \pqty{0, -1},\quad
\bm{q}_\text{tr} = k_\theta \pqty{\frac{\sqrt{3}}{2}, \half},\quad
\bm{q}_\text{tl} = k_\theta \pqty{\frac{-\sqrt{3}}{2}, \half}, \nn
\end{equation}
which are characterized by matrices,
\begin{equation}
T_\text{b} = \begin{bmatrix}1 & 1 \\ 1 & 1\end{bmatrix},\quad
T_\text{tr} = \begin{bmatrix}e^{-i\phi} & 1 \\ e^{i\phi} & e^{-i\phi}\end{bmatrix},\quad
T_\text{tl} = \begin{bmatrix}e^{i\phi} & 1 \\ e^{-i\phi} &
e^{i\phi}\end{bmatrix}
\end{equation}
where $\phi = 2\pi/3$ \cite{bistritzer11a,bistritzer11b,degail11}. Repeated
hopping generates a honeycomb lattice in the momentum space. Truncating the
continuum model Hamiltonian \cite{dossantos07,bistritzer11a,dossantos12}
at the first honeycomb shell, gives rise to the following eight-band
Hamiltonian:
\begin{equation}
\mathcal{H}_\theta
= \begin{bmatrix}
h_{\theta/2}(\bm{k}) & wT_\text{b} & wT_\text{tr} & wT_\text{tl} \\
wT^\dagger_\text{b} & h_{-\theta/2}(\bm{k}_\text{b}) & 0 & 0 \\
wT^\dagger_\text{tr} & 0 & h_{-\theta/2}(\bm{k}_\text{tr}) & 0 \\
wT^\dagger_\text{tl} & 0 & 0 & h_{-\theta/2}(\bm{k}_\text{tl})
\end{bmatrix},
\label{e:eight-band}
\end{equation}
where $\bm{k}_j = \bm{k} + \bm{q}_j$ and $w \approx \SI{110}{meV}$ is the
interlayer hopping energy \cite{bistritzer11a,python19}.
The Hamiltonian $\mathcal{H}_\theta$ acts on a four-dimensional vector of
two-component spinors, which is why it is called an eight-band model
\cite{python19}.

\begin{figure*}[ht]
\centering
\includegraphics[width=0.32\textwidth]{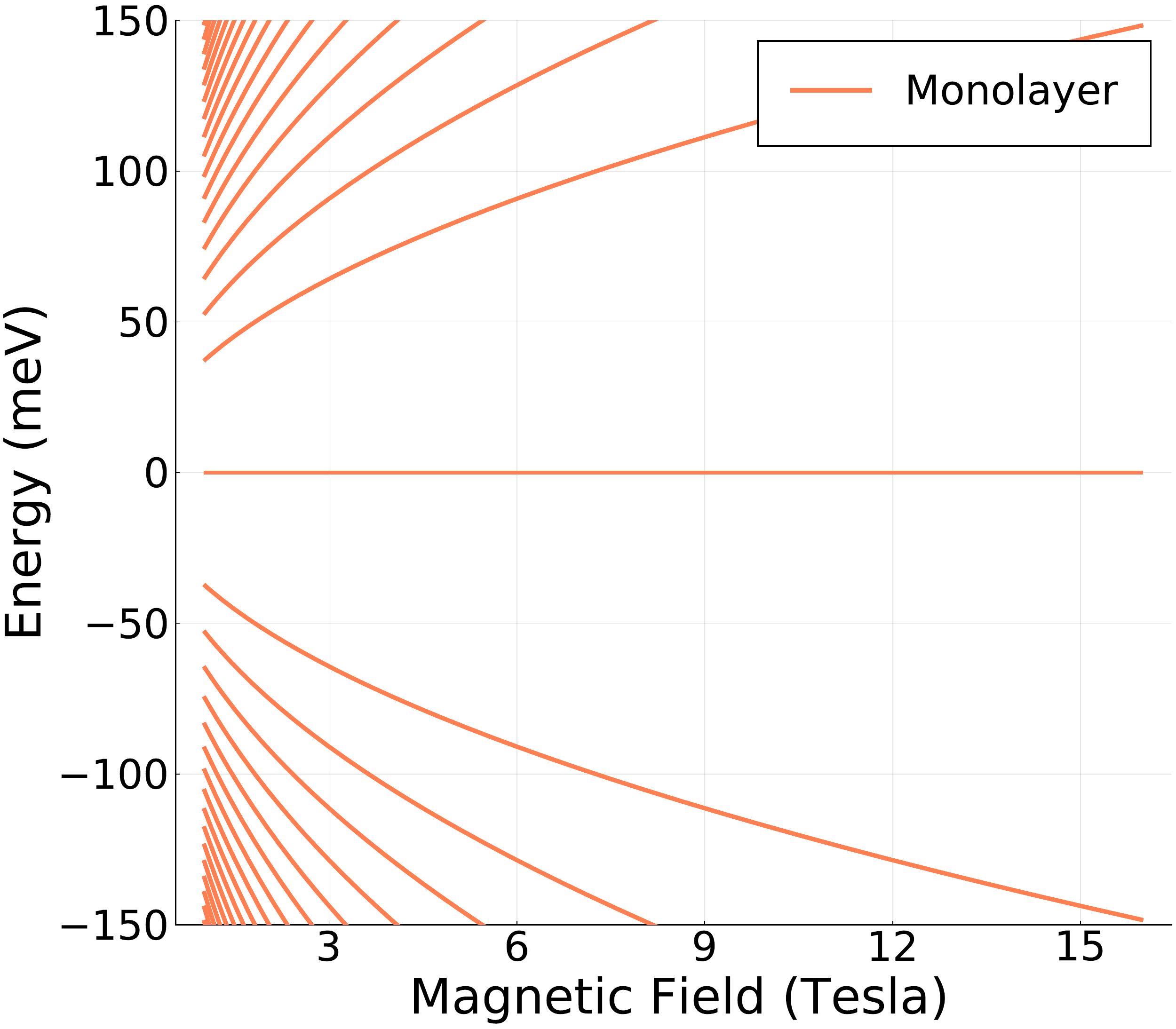}
\includegraphics[width=0.32\textwidth]{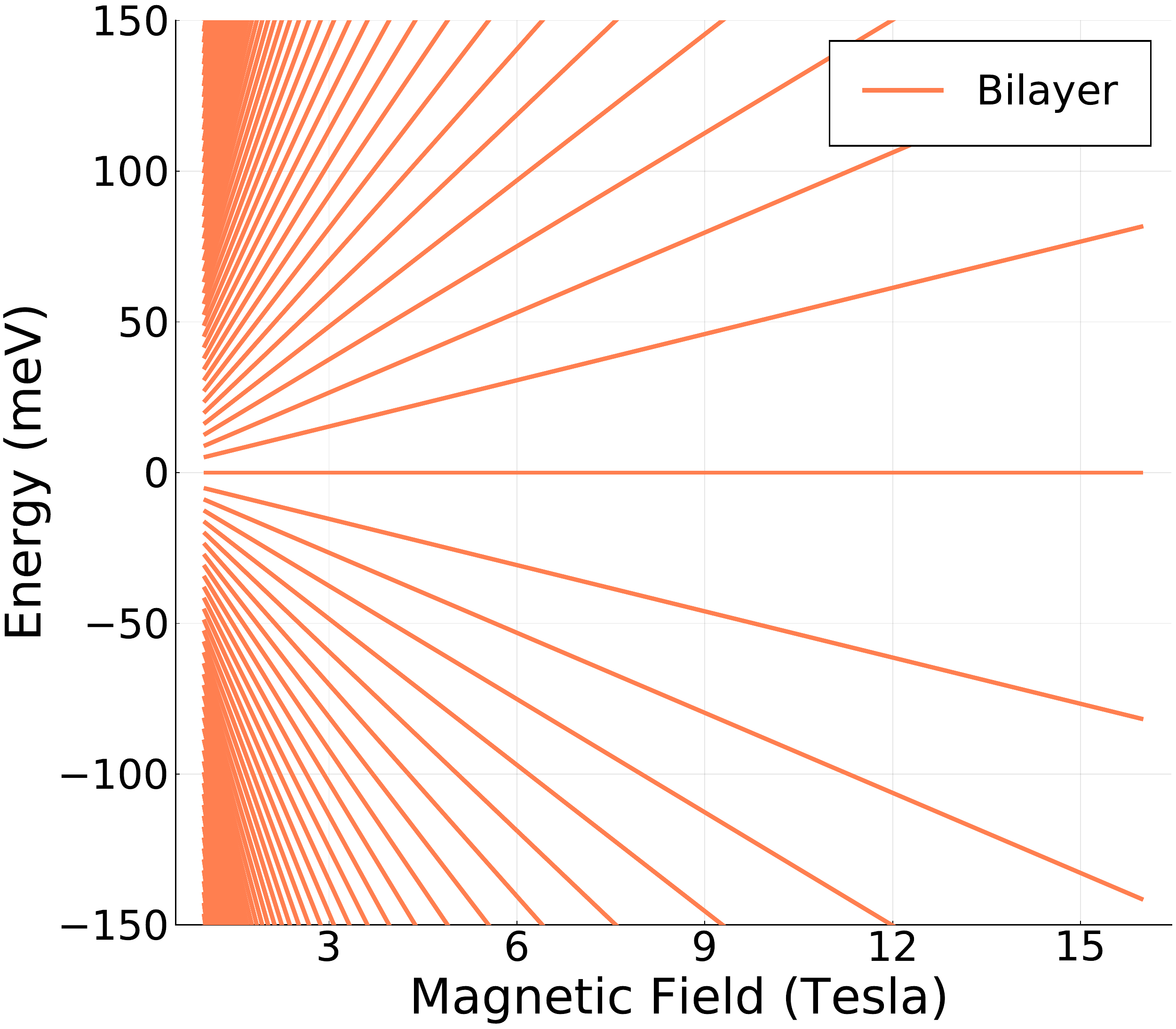}

\includegraphics[width=0.32\textwidth]{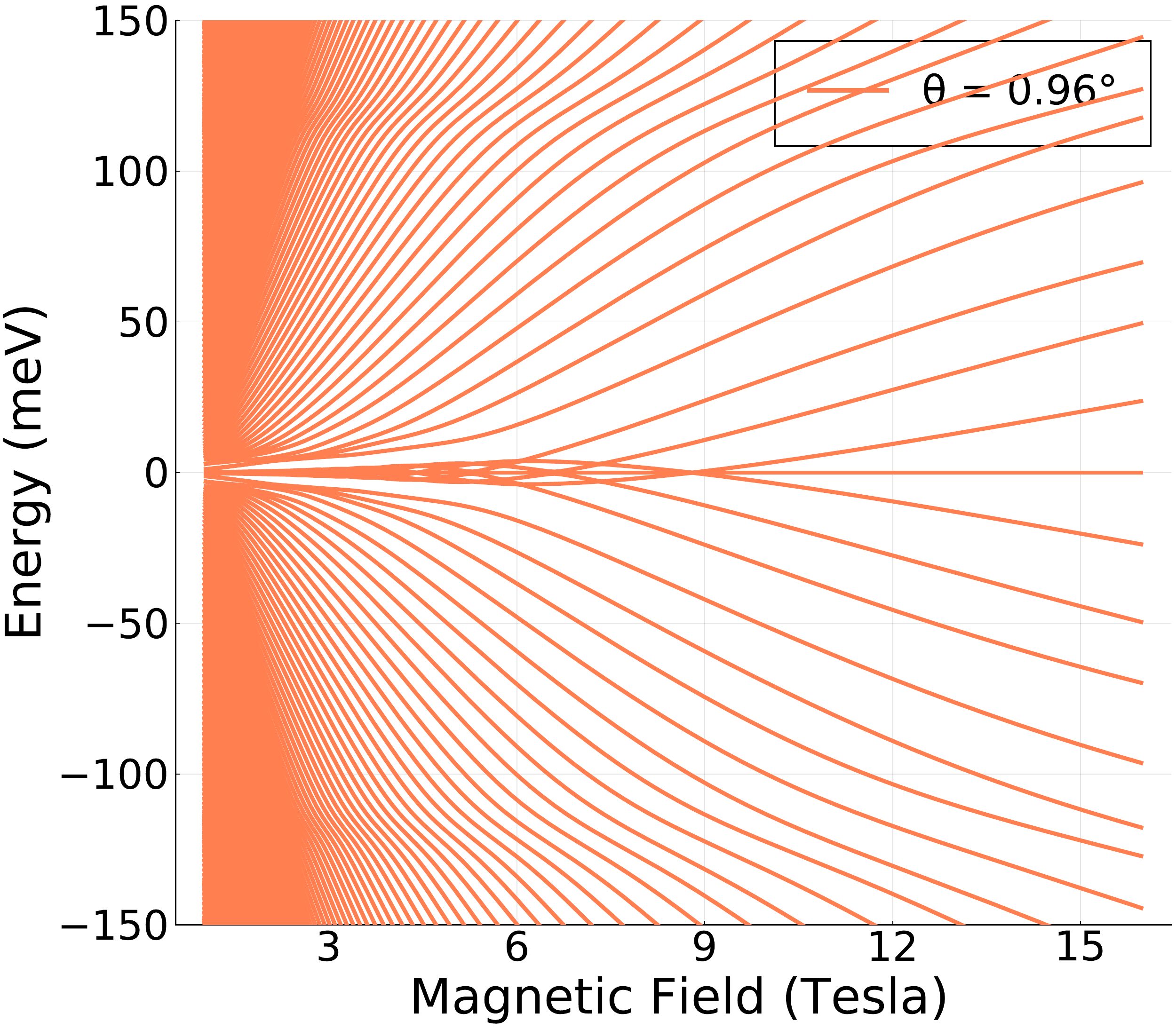}
\includegraphics[width=0.32\textwidth]{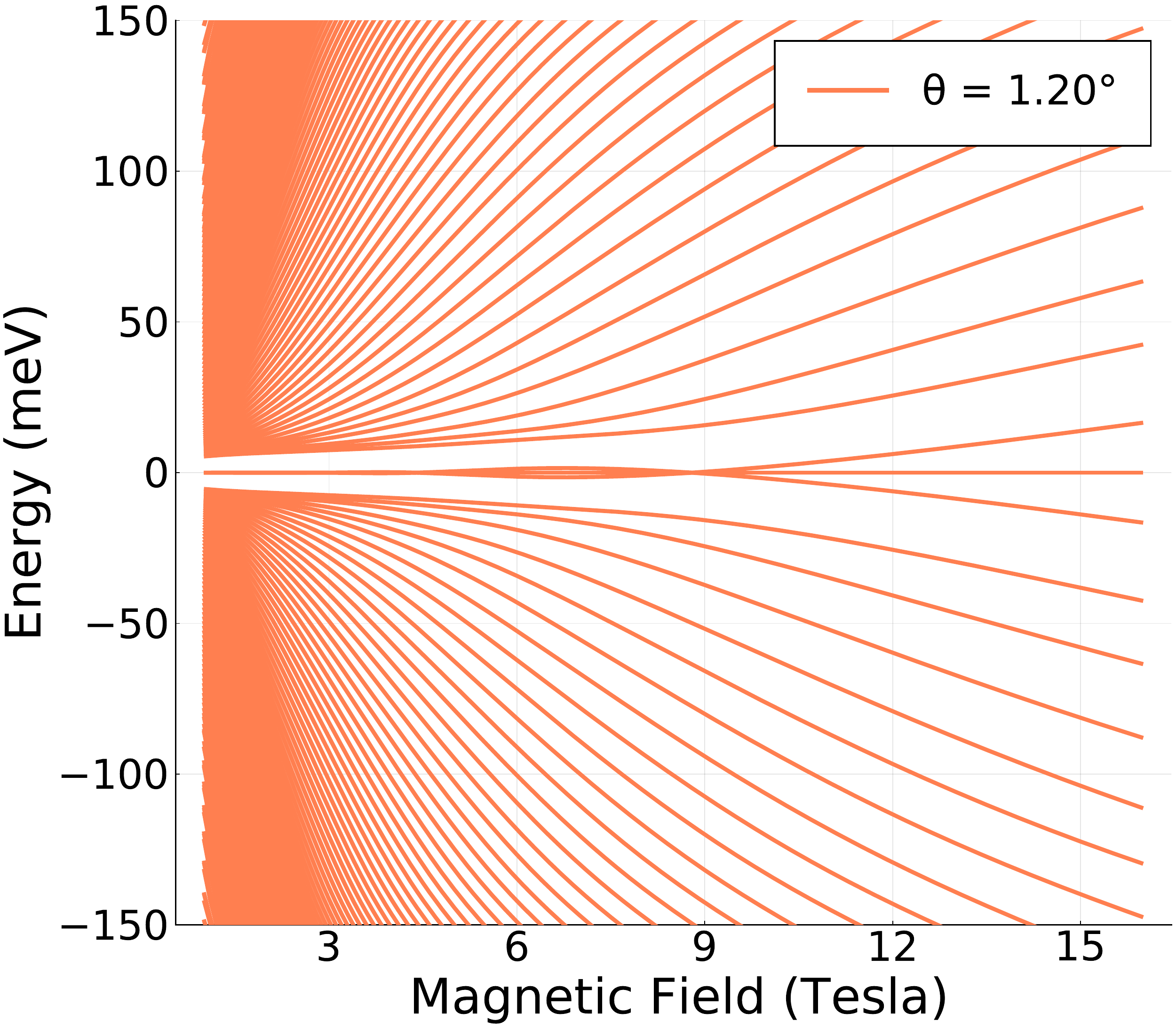}
\includegraphics[width=0.32\textwidth]{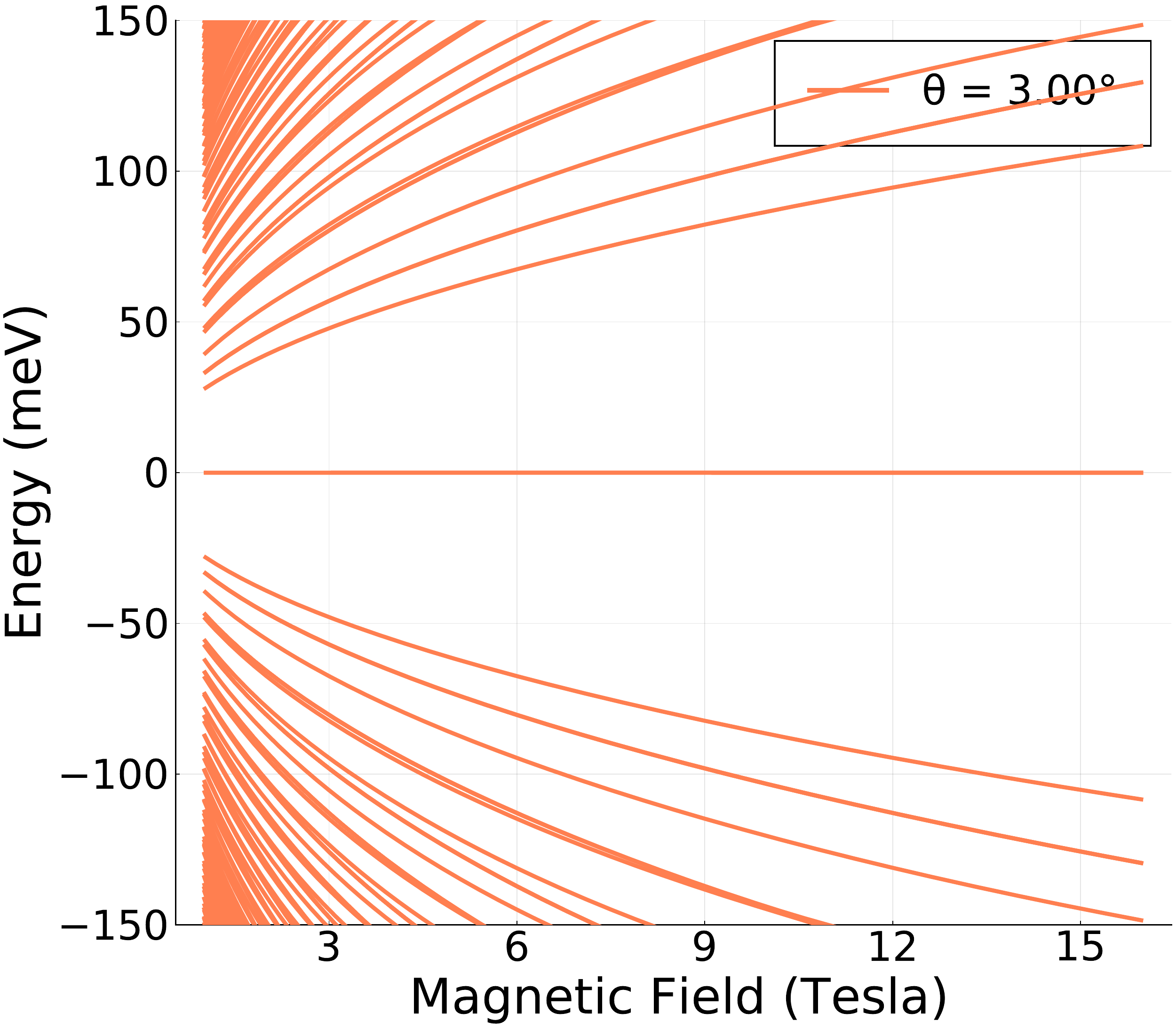}

\caption{\label{fig:fan-plots} Landau level spectra for twisted bilayer
graphene. The most important feature of these plots is the dispersion
of energy levels concerning the magnetic field. For monolayer
Landau levels go as $E_n \sim \sqrt{B}$, for bilayer the dispersion is
linear $E_n \sim B$. With a small twist, Landau levels around zero
energy start having a flat dispersion, and as the twist is increased the
Landau level spectrum starts looking remarkably similar to the
monolayer. The peculiar nature of magic-angle induced flat bands near zero
energy for $\theta^\star = \ang{0.96}$ should be noted.}
\end{figure*}

The angle dependence of $h_\theta$ is parametrically small and can be
neglected, and it was shown in~\cite{bistritzer11a} that the eight-band
approximation reproduces correct Fermi velocity with reasonable accuracy
to the first magic angle. Up to a scale factor, the electronic structure of
the eight-band model depends on dimensionless parameter $\alpha = w/\hbar
v_F k_\theta$, in terms of which we can write renormalized Fermi velocity
\cite{bistritzer11a},
\begin{equation}
\frac{v_F'}{v_F} = \frac{1 - 3\alpha^2}{1 + 6\alpha^2},
\end{equation}
which vanishes at $\theta^\star \approx \ang{0.96}$. It is where the magic angle
occurs in the model. Moreover, it is the only magic angle
the eight-band model can reproduce since we are truncating the momentum space
lattice at the first honeycomb shell. Refs \cite{bistritzer11b,moon12,zhang19}
describe, in detail, the procedure for obtaining Landau levels in TBG, with
full continuum model Hamiltonian. We also note that several qualitative
features of TBG like flat bands and interpolation of electronic
the structure between bilayer and monolayer behavior can also be seen in ab-initio
tight-binding calculations of the band structure \cite{morell10}.

\subsection{Numerically computing Landau levels in twisted bilayer graphene}

As the ladder operators introduced in the previous section obey $[\Pi,
\Pi^\dagger] = 1$, we have harmonic oscillator states $\ket{0}, \ket{1},
\ket{2},\ldots$ satisfying $\Pi^\dagger \Pi \ket{n} = n\ket{n}$.
These states constitute a complete, orthonormal basis set for this Hilbert
space.
$\Pi^\dagger$ and $\Pi$ act as raising and lowering operators for these
states with $\Pi^\dagger \ket{n} = \sqrt{n + 1}\ket{n + 1}$ and
$\Pi \ket{n} = \sqrt{n}\ket{n - 1}$ respectively. Using these relations, it can
be verified that the matrix representation of ladder operators in this basis is
\begin{equation}
\mel{n}{\Pi}{m} = \sqrt{m} \delta_{n,m-1}\quad
\text{and}\quad
\mel{n}{\Pi^\dagger}{m} = \sqrt{m + 1} \delta_{n,m+1}.
\end{equation}

To find Landau level spectrum, the substitution $\hbar
\bm{k} \to \bm{\pi}$ is made in an eight-band Hamiltonian, and it is written in terms of these ladder operators.
However, as the harmonic oscillator states do not constitute an eigenbasis of the Hamiltonian, this representation is not diagonal, and the energy eigenvalues have to be determined numerically. \cite{code}.

In principle, the basis $\Bqty{\ket{n}}_{n = 0, 1, 2,\ldots}$, is
infinite, but for practical purposes, we truncate it after a large,
but finite number of states:~$\ket{0}, \ket{1}, \ldots,\ket{N}$. For all
calculations, we have retained $N = 500$ harmonic oscillator states for finding
Landau levels. In our trials, it was observed that retaining fewer Landau
levels resulted in deviations in energy eigenvalues at low magnetic fields,
while retaining more than $500$ Landau levels resulted in considerable execution times without any significant improvement in  accuracy.

Landau level spectra obtained by numerically diagonalizing the Hamiltonian have been plotted in Fig.~\ref{fig:fan-plots}. The most striking feature of these
plots are the dispersion of energies to the magnetic field.
In TBG, especially at magic angles, the peculiar nature of flat bands near zero
energy should be noted.
At larger twist angles, as the two layers get decoupled, qualitative features
of the TBG spectrum and the dispersion for the magnetic field are very
similar to monolayer, except for a renormalized Fermi velocity
\cite{bistritzer11a, python19}.
The efficiency of the Otto cycle depends,
almost exclusively, on the dispersion of Landau levels for
the magnetic field. It is discussed in detail in the sections dealing with quantum heat engine cycles and results and discussion.
\section{Quantum heat engine cycle}

For the heat engine, we shall consider an ensemble of single electron states in the
conduction band \cite{munoz14, pena15, pena19, pena20}. We take Landau levels
$\ket{\psi_n}$ with energies $E_n$ and occupation probabilities $p_n$, so that
the density matrix is $\rho = \sum_n p_n \dyad{\psi_n(B)}$. Average energy of this
ensemble, $U = \Tr(\rho \mathcal{H}_\theta) = \sum_n p_n E_n$
is identified as \emph{internal energy} of the system, and we can state a
quantum version of the first law of thermodynamics \cite{kieu04, quan05, quan07},
\begin{equation}
\dd U = \dd Q + \dd W = \sum_n E_n \dd p_n + \sum_n p_n \dd E_n.
\label{e:quantum-flot}
\end{equation}
Since thermodynamic entropy $S = -k_B \sum_n p_n \ln p_n$, and in classical
thermodynamics, heat exchanged $\dd Q = T \dd S$, we identify $\dd Q =
\sum_n E_n \dd p_n$ and work as $\dd W = \sum_n p_n \dd
E_n$~\cite{kieu04,quan05,quan07}.
From density matrix, we can calculate von Neumann entropy,
\begin{equation}
S(T, B) = -k_B \Tr(\rho \ln \rho) = -k_B \sum_n p_n \ln p_n.
\end{equation}

Occupation probabilities of different energy levels are determined by
the temperature of working substance \cite{quan07,munoz14}. The temperature of the working substance is controlled either by keeping it in equilibrium with a heat bath
and varying its temperature quasistatically, or by coupling it
to hot and cold reservoirs alternatively \cite{abah12, rossnagel14,
vinjanampathy15}. At temperature $T$, occupation probabilities satisfy
The Boltzmann distribution, which is given as,
\begin{equation}
p_n(T, B) = \frac{e^{-\beta E_n(B)}}{Z(T, B)};\quad
Z(T, B) = \sum_{n=0}^\infty e^{-\beta E_n(B)},
\label{e:boltz-dist}
\end{equation}
with $\beta = 1/k_B T$ and $Z(T, B)$ being partition function.
In what follows, we shall assume that the thermal reservoir is a classical
object and that its temperature can be varied quasistatically. We shall also
assume that external magnetic field can be varied quasistatically to modulate
Landau levels $E_n$, and their occupation probabilities $p_n$.

\begin{figure}[ht]
\centering
\includegraphics[width=0.36\textwidth]{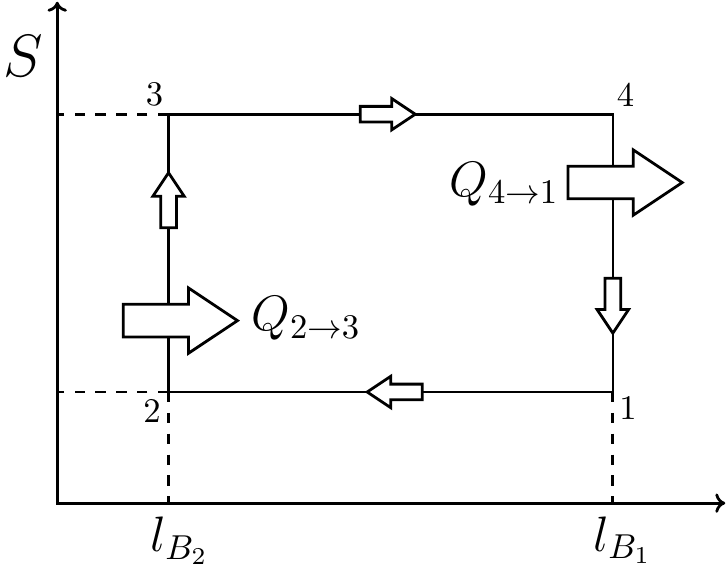}
\caption{\label{fig:otto-cycle}
Four strokes of Otto cycle on an
entropy--magnetic field plot. The cycle starts with an adiabatic
compression $1\to 2$, in which Landau radius decreases due to an
increase in magnetic field, followed by an isochoric absorption of heat
$2\to 3$. Next, magnetic field is decreased adiabatically $3\to 4$
and finally heat is rejected isochorically $4\to 1$ to return the
system to its initial state.}
\end{figure}

Quantum Otto cycle consists of four strokes, operating between magnetic
field strengths $B_1$ and $B_2$ (with $B_2 > B_1$), and temperatures $T_C$ and
$T_H$ (with $T_H > T_C$). In order to draw parallels with classical Otto
cycle, it is easier to state the compression and expansion strokes in terms of
decreasing and increasing \emph{Landau radius} $l_B = \sqrt{\hbar / eB}$.

Before discussing details of the cycle, we take a moment to note
the differences between the general and strict versions of the adiabatic stroke. In
the general version of the adiabatic stroke, the working substance is kept in thermal
equilibrium, and the temperature of the heat reservoir is changed
gradually. The entropy remains constant throughout the process. In
particular, in an adiabatic process when temperatures and magnetic fields
change $(T_i, B_i) \to (T_f, B_f)$, we require $S(T_i, B_i) = S(T_f, B_f)$. In
the strict version of the adiabatic process, we impose a stronger constraint and require the
occupation probabilities of energy states to remain unchanged $p_n(i) =
p_n(f)$~\cite{quan07,pena20b}. In particular, at the end of the stricter version of the adiabatic
stroke, the system need not be in a state with well-defined
temperature~\cite{quan07}.  The "stricter" and "general" versions of the adiabatic stroke might look very different and lead to different work output and efficiency. However, both these conditions can be shown to be equivalent when the energy levels change in the same ratio (cf. Appendix B). In what follows, we focus on a heat engine cycle with general adiabatic stroke and leave the details of the cycle with strict adiabatic strokes to appendix B.

The first stroke is an adiabatic compression in which Landau radius is
reduced by gradually increasing the external magnetic field. Due to this
changing magnetic field, we have $l_{B_1} \to l_{B_2}$.
However, since entropy has to be held constant, temperature must also change $T_C \to
T_2$, to satisfy adiabatic condition $\Delta S = 0$ and
the intermediate temperature $T_2$ is determined by the condition
\begin{equation}
S(T_C, B_1) = S(T_2, B_2).
\label{e:adiabatic-1}
\end{equation}

In the second stroke, the working substance absorbs heat from the reservoir,
while Landau radius is held constant at $l_{B_2}$ and acquires a temperature
$T_H$ at the end of the process.
This is called a \emph{hot isochore}~\cite{abah12,rossnagel14}. Heat
absorbed in this stroke can be calculated from (\ref{e:quantum-flot}),
\begin{align}
Q_{2\to 3} & = \int_2^3 \sum_{n=0}^\infty E_n(B_2) \dd p_n \nn \\
& = \sum_{n=0}^\infty E_n(B_2)[p_n(T_H, B_2) - p_n(T_2, B_2)]
\label{e:q-in}
\end{align}
The next stroke is an adiabatic expansion and involves an increase of Landau radius $l_{B_2} \to l_{B_1}$. As the temperature changes $T_H \to T_4$ the general adiabatic condition reads:
\begin{equation}
S(T_H, B_2) = S(T_4, B_1).
\label{e:adiabatic-2}
\end{equation}

In the final stroke, heat is lost
to reservoir, with Landau radius being held constant at $l_{B_1}$ as the system
attains the temperature $T_C$ and the cycle can be started over again. This
process is called a \emph{cold isochore}~\cite{abah12,rossnagel14}. Heat
exchanged in this stroke can be calculated as before,
\begin{equation}
Q_{4\to 1} = \sum_{n = 0}^\infty E_n(B_1)[p_n(T_C, B_1) - p_n(T_4, B_1)].
\label{e:q-out}
\end{equation}

Since no heat exchange occurs in adiabatic processes, and working
substance returns to its initial state at end of cycle, we can use
quantum first law with $\Delta U = 0$ to write work output of
engine as
\begin{equation}
\vqty{W_\text{O}} = Q_\text{cycle} = \vqty{Q_{2\to 3}} - \vqty{Q_{4\to 1}},
\label{e:otto-work}
\end{equation}
while efficiency is given by,
\begin{align}
\eta_\text{O} & = \vqty{\frac{W_\text{O}}{Q_\text{in}}} = \frac{\vqty{Q_{2\to 3}}
- \vqty{Q_{4\to 1}}}{\vqty{Q_{2\to 3}}}
= 1 - \vqty{\frac{Q_{4\to 1}}{Q_{2\to 3}}} \\
& = 1 - \vqty{\frac{\sum_n E_n(B_1)[p_n(T_C, B_1) - p_n(T_4,
B_1)]}{\sum_n E_n(B_2)[p_n(T_H, B_2) - p_n(T_2, B_2)]}},
\label{e:otto-efficiency}
\end{align}
where in the last line, we used (\ref{e:q-in}) and (\ref{e:q-out}).

The discussion up to this point has been entirely general since all these
results are direct consequences of the quantum first law of thermodynamics.
Eqs.~(\ref{e:otto-work}) and (\ref{e:otto-efficiency}) are equally applicable
to any quantum working substance coupled to a classical thermal reservoir. Very
similar expressions for work and efficiency appear, for example, in \cite{quan05,
quan07,munoz14,pena15,pena19}.

\section{Results and discussion}

In case of $AB$ stacked bilayer graphene, where an analytical expression
for Landau levels is known i.e.
\begin{equation}
E_n = \hbar \omega_B \sqrt{n (n - 1)},\quad n = 0, 1, 2,\ldots
\end{equation}
with $\omega_B = eB/m_\text{eff}$ \cite{mccann06, mccann13}, we note that the
energy levels change in the same ratio and constraints on $T_2$ and $T_4$ are
equivalent to the strict adiabatic conditions (cf. Appendix~B)
\begin{equation}
p_n(T_C, B_1) = p_n(T_2, B_2)
\text{ and }
p_n(T_H, B_2) = p_n(T_4, B_1).
\end{equation}
We can use (\ref{e:otto-efficiency}) to derive the Otto efficiency,
\begin{align}
\eta_\text{O}^\text{bi} = 1 - \vqty{\frac{\omega_{B_1}}{\omega_{B_2}}} = 1 -
\pqty{\frac{l_{B_2}}{l_{B_1}}}^2 = 1 - r_C^{-2},
\end{align}
where we have defined the \emph{compression ratio} $r_C = l_{B_1}/l_{B_2}$ so
that the efficiency is reminiscent of the classical expression $\eta_O = 1 -
r_C^{-(\gamma - 1)}$ \footnote{$\gamma = C_p/C_V$ is the ratio of specific
heats at constant pressure and constant volume}.
Similarly, for monolayer graphene, we have \cite{goerbig11}
\begin{equation}
E_n = \frac{\hbar v_F}{l_B} \sqrt {2n},\quad n = 0, 1, 2,\ldots
\end{equation}
and therefore,
\begin{equation}
\eta_\text{O}^m = 1 - \vqty{\frac{l_{B_2}}{l_{B_1}}} = 1 - r_C\inv.
\end{equation}
As already stated, in both monolayer and bilayer graphene,
general and stricter versions are equivalent.

\begin{figure}[htb]
\includegraphics[width=0.36\textwidth]{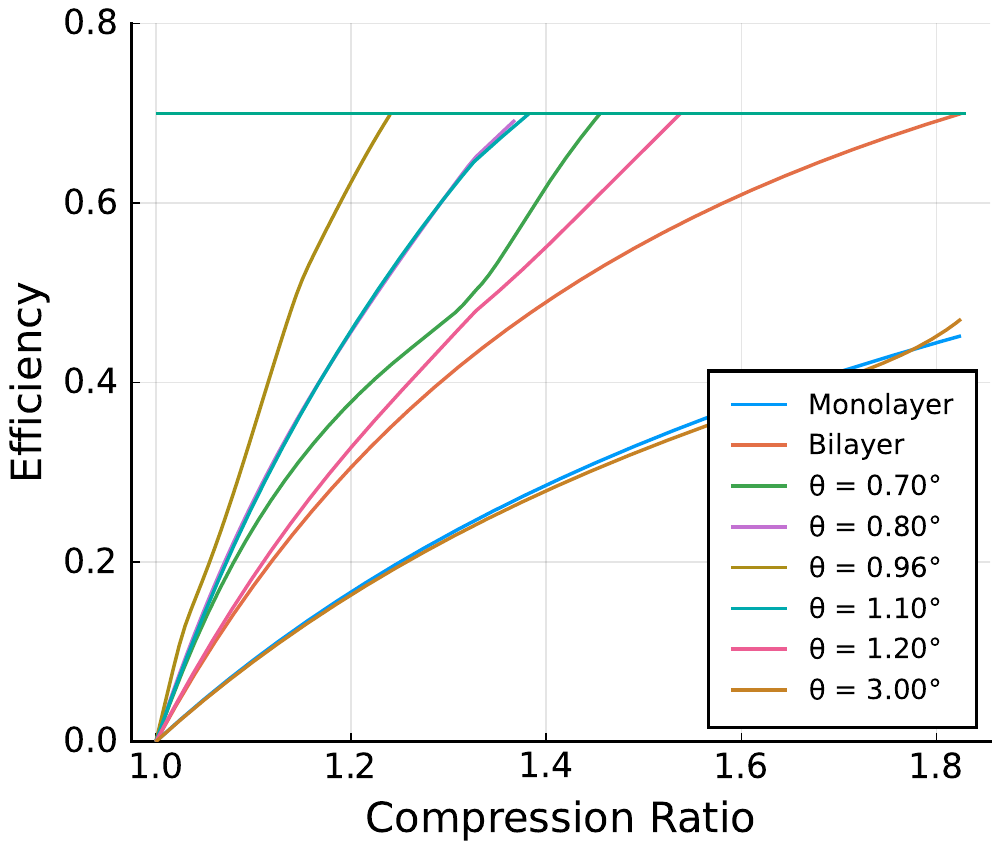}
\caption{\label{fig:otto-efficiency} Efficiencies as a function of the
compression ratio for Otto cycles operating between $T_C = \SI{30}{K}$ and
$T_H = \SI{100}{K}$, and $B_1 = \SI{5.0}{T}$. Different twist angles are plotted together for comparison.}
\end{figure}

\begin{figure}[htb]
\includegraphics[width=0.36\textwidth]{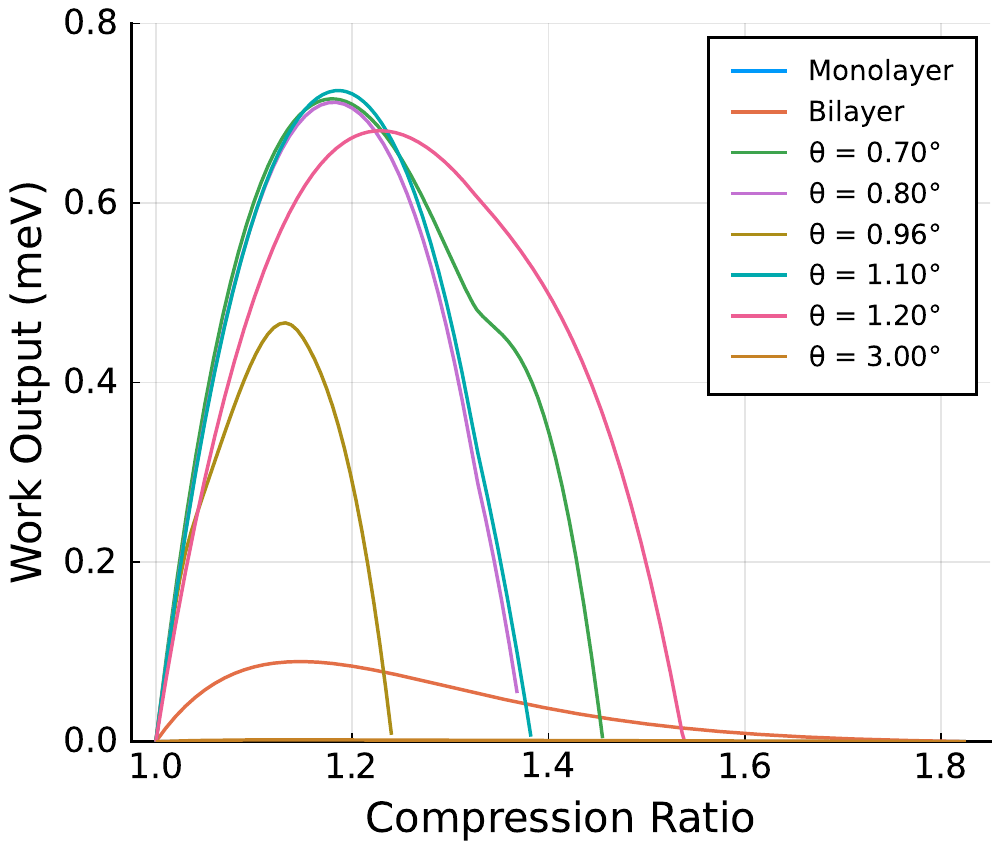}
\caption{\label{fig:otto-work} Work output as a function of
compression ratio at different twist angles, for cycles operating
between $T_C = \SI{30}{K}$ and $T_H = \SI{100}{K}$, and magnitude $B_1 =
\SI{5.0}{T}$.
Different twist angles are plotted together for comparison.}
\end{figure}

For twisted bilayer graphene, a simple expression for Landau level
energies is not known, and therefore an analytic expression for efficiency
cannot be derived. In the classical cycle temperatures $T_2$ and $T_4$
have to be determined numerically from adiabatic conditions
(\ref{e:adiabatic-1}) and (\ref{e:adiabatic-2}), and efficiency has to be
computed directly from (\ref{e:otto-efficiency}). Efficiencies and work outputs
are plotted for different angles in Fig.~\ref{fig:otto-efficiency} and
Fig.~\ref{fig:otto-work} respectively.
For all numerical computations, $N = 500$ Landau levels were retained.

Some partial insight into increased efficiencies can be ascertained by looking
at qualitative differences in Landau level plots for monolayer, bilayer, and
magic angle twisted bilayer graphene. If the Landau levels have the dispersion
$E_n(B) = l_B^{-\alpha} f(n) = (eB/\hbar)^{\alpha/2} f(n)$ then, from
(\ref{e:otto-efficiency}) we have
\begin{equation}
\eta_\text{O} = 1 - r_C^{-\alpha}.
\label{e:gen-efficiency}
\end{equation}
Furthermore, efficiencies obtained by direct numerical computation can be
fitted for the parameter $\alpha$ in (\ref{e:gen-efficiency}) and
a larger value of $\alpha$ in dispersion of $E_n$ is responsible for
higher efficiency. It is what we see in Landau level plots for
twisted bilayer.
After attaining a maximum at $\theta^\star = \ang{0.96}$ (magic angle) the
efficiency starts falling for larger twist angles until it coincides with
monolayer efficiency for $\theta = 3.0\dg$. It is to be expected because, for larger twists, the two layers get decoupled.

The take-home message of our paper is the following: proposed quantum Otto
engine has the highest efficiency at magic angle $\theta^\star = 0.96\dg$.

Work output is obtained by directly computing the difference
between heat absorbed (\ref{e:q-in}) and heat lost (\ref{e:q-out}) during the
cycle. Work is plotted as a function of compression ratio in
Fig.~\ref{fig:otto-work}. For each case, work output initially increases
as the compression ratio is increased, and after attaining a maximum, starts
falling and eventually reaches zero just as efficiency reaches Carnot
limit $\eta_\text{C} = 1 - T_C/T_H$. Zero work output at
Carnot efficiency can be interpreted as a manifestation of the
second law of thermodynamics in these systems. In particular, we note that the
proposed heat engine cannot surpass the Carnot limit despite operating with a
quantum working substance as the cycle is composed of equilibrium processes
committed to operating between two temperatures
\cite{munoz14,niedenzu15,vinjanampathy15}.

Finally, we note that a dip in work output at the magic angle. Due to flat bands,
Landau levels which are predominantly occupied during the cycle (according to
Eq. \ref{e:boltz-dist}), have energy very close to zero. Therefore heat absorbed
(\ref{e:q-in}) and rejected (\ref{e:q-out}) during the cycle are significantly
smaller, resulting in lower work output.

\begin{table}[htp]
\caption{\label{table}A comparison of efficiencies and work outputs for Otto cycle.}
\begin{ruledtabular}
\bgroup
\def\arraystretch{1.5}
\begin{tabular}{cl|c d}
\multicolumn{2}{c|}{Working Substance} & Efficiency & \multicolumn{1}{r}{Work (meV)} \\
\hline
\multicolumn{2}{c|}{Monolayer Graphene} & $1 - r_C^{-1.00}$ & 0.000169 \\
\multicolumn{2}{c|}{Semiconductor} & $1 - r_C^{-1.98}$ & 1.143 \\
\multicolumn{2}{c|}{Bilayer Graphene} & $1 - r_C^{-2.00}$ & 0.0758 \\
& $\theta = 0.70\dg$ & $1 - r_C^{-2.68}$ & 0.669 \\
& $\theta = 0.80\dg$ & $1 - r_C^{-3.48}$ & 0.650 \\
Twisted Bilayer & $\theta^\star = \ang{0.96}$ & $1 - r_C^{-5.03}$ & 0.438 \\
Graphene & $\theta = 1.10\dg$ & $1 - r_C^{-3.47}$ & 0.671 \\
& $\theta = 1.20\dg$ & $1 - r_C^{-2.37}$ & 0.608 \\
& $\theta = 3.00\dg$ & $1 - r_C^{-0.98}$ & 0.00159
\end{tabular}
\egroup
\end{ruledtabular}
\end{table}

\section{Experimental realization and conclusion}

\begin{figure}[htb]
\includegraphics[width=0.36\textwidth]{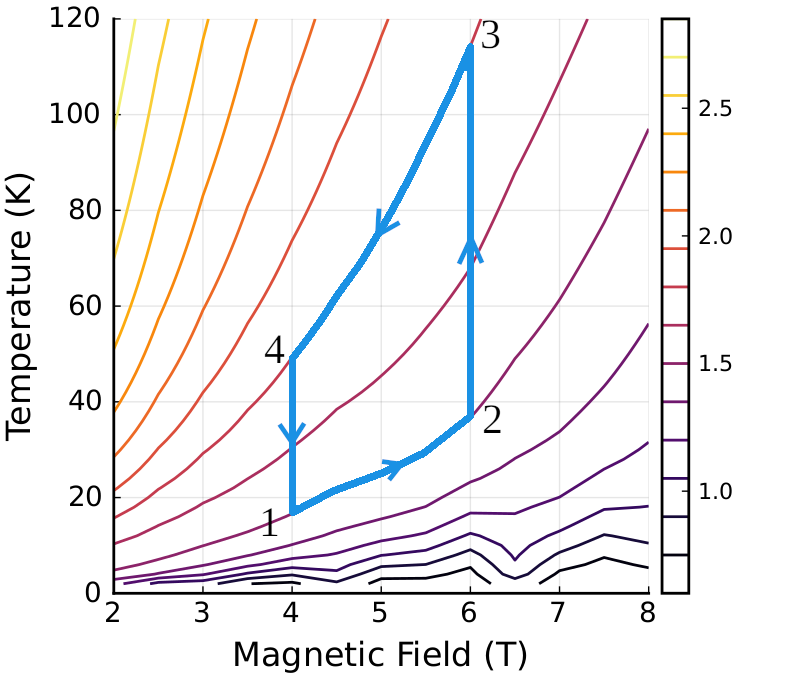}
\caption{\label{fig:entropy} Isentropic lines on the temperature--magnetic
field plane for twisted bilayer graphene at magic angle. A possible
Otto cycle with general adiabatic conditions is traced out in blue.}
\end{figure}

Until this point, we have not referred to any specific properties of the heat
reservoir to which the working substance couples during the hot and cold
isochores. In experimental realizations of quantum Otto cycles, the details of
the heat reservoir depend strongly on the working substance. For example, in a
single-ion heat engine, \cite{abah12} the hot and cold reservoirs are realized
by detuned lasers at different frequencies, while in a spin quantum heat
engine \cite{peterson19} it is implemented by a suitable sequence of rf pulses.
Recently, there have been experiments measuring the magnetic entropy of magic-angle graphene at low temperatures, which has led to the discovery of an
electronic phase transition at zero magnetic field \cite{saito21, rozen21}.
These experiments demonstrate the possibility of precisely controlling the thermodynamic state of MATBG in a laboratory that could implement
isochores of the Otto cycle proposed in this paper. In particular,
Fig.~\ref{fig:entropy} shows how the temperature and magnetic field should be simultaneously changed to implement the cycle.

We note that a Carnot cycle with MATBG can be designed analogously by
using two adiabatic and two isothermal strokes. While the efficiency of a
Carnot cycle is fixed at $\eta_C = 1 - T_C/T_H$ independent of expansion ratio \cite{quan07, munoz14}, it will be interesting to see what impact magic angle twists have on work output. After looking at a highly efficient quantum heat engine based on MATBG, a natural question to ask next is if it might be possible to use MATBG to design a nanoscale refrigerator with a high coefficient of performance. Both these possibilities will be explored in a future work \cite{future}.
\acknowledgments
This work was supported by the grants: 1. Josephson junctions with strained Dirac materials and their application in quantum information processing, Science \& Engineering Research Board (SERB) Grant No. CRG/20l9/006258, and 2. Nash equilibrium versus Pareto optimality in N-Player games, Science \& Engineering Research Board (SERB) MATRICS Grant No. MTR/2018/000070.
\bibliography{twist-pr}
\appendix

\section{Equivalence of conservation of thermal populations and the adiabatic conditon}

For monolayer graphene we have
\begin{equation}
E^\text{mono}_n(B) = \frac{\hbar v_F}{l_B}\sqrt{2n},
\end{equation}
where $l_B = \sqrt{\hbar/eB}$.
If the magnetic field is changed from $B_1$ to $B_2$, we get
\begin{align}
E^\text{mono}_n(B_2) & = \frac{\hbar v_F}{l_{B_2}}\sqrt{2n} \nn \\
& = \sqrt{\frac{B_2}{B_1}} \frac{\hbar v_F}{l_{B_1}}\sqrt{2n}
= \sqrt{\frac{B_2}{B_1}} E^\text{mono}_n (B_1).
\end{align}
Similarly, as for bilayer graphene the Landau levels are $E^\text{bi}_n = \hbar \omega_B
\sqrt{n(n-1)}$, with $\omega_B = eB/m_\text{eff}$, we get
\begin{align}
E^\text{bi}_n(B_2) & = \hbar \omega_{B_2} \sqrt{n(n-1)} \nn \\
& = \frac{B_2}{B_1} \omega_{B_1} \sqrt{n(n-1)}
= \frac{B_2}{B_1} E^\text{bi}_n (B_1).
\end{align}
In both these cases, we have $E_n(B_2) = \zeta E_n(B_1)$, where $\zeta$ is a
constant independent of $n$, but which depends on the ratio $B_2/B_1$.

First we note that the conservation of thermal populations $P_n(B_1, T_1) =
P_n(B_2, T_2)$ implies $S(B_1, T_1) = S(B_2, T_2)$ because $S(B, T) = \sum_n
P_n(B, T) \ln P_n(B, T)$. The converse can be proven if we consider the special
case when energy levels change in the same ratio, i.e., $E_n(B) \to \zeta
E_n(B)$.
In thermal equilibrium, the occupation probability of each Landau level $E_n$
is given by the Boltzmann
distribution
\begin{equation}
P_n(B, T) = \frac{e^{-\beta E_n(B)}}{Z(B, T)};
\quad
Z(B, T) = \sum_{n=0}^\infty	e^{-\beta E_n(B)},
\end{equation}
and we have
\begin{equation}
\frac{P_n(B_1, T_1)}{P_m(B_1, T_1)}
= \frac{e^{-\beta_1 E_n(B_1)}}{e^{-\beta_1 E_m(B_1)}}
\end{equation}
and
\begin{equation}
\frac{P_n(B_2, T_2)}{P_m(B_2, T_2)}
= \frac{e^{-\beta_2 E_n(B_2)}}{e^{-\beta_2 E_m(B_2)}}
= \frac{e^{-\zeta \beta_2 E_n(B_1)}}{e^{-\zeta \beta_2 E_m(B_1)}}
\end{equation}
If the temperature at the end of the cycle is chosen to be $T_2 = \zeta T_1$ so
that $\beta_1 = \zeta \beta_2$, then for every $m$ and $n$ we have
\begin{align}
& \frac{P_n(B_1, T_1)}{P_m(B_1, T_1)} = \frac{P_n(B_2, T_2)}{P_m(B_2, T_2)}
\nn \\
\implies & \frac{P_n(B_1, T_1)}{P_n(B_2, T_2)} = \frac{P_m(B_1,
T_1)}{P_m(B_2, T_2)} = \lambda \quad \text{(say)}.
\label{e:prob-ratio}
\end{align}
Now, we if impose the adiabatic condition $S(B_1, T_1) = S(B_2, T_2)$, we get
\begin{align}
\sum_n & P_n (B_2, T_2) \ln P_n (B_2, T_2) \nn \\
& = \sum_n P_n (B_1, T_1) \ln P_n (B_1, T_1) \nn \\
& = \sum_n \lambda P_n (B_2, T_2) [\ln \lambda + \ln P_n(B_2, T_2)] \nn \\
& = \lambda \sum_n P_n (B_2, T_2) \ln P_n (B_2, T_2) + \lambda \ln
\lambda,
\end{align}
which gives the following equation for $\lambda$
\begin{align}
(\lambda - 1) S(B_2, T_2) = \lambda\ln\lambda.
\end{align}
Since (\ref{e:prob-ratio}) holds independent of particular values of
temperature and magnetic field, the solution to $\lambda$ in the above equation
must not depend on $S(B_2, T_2)$. In such a scenario, the only
solution is $\lambda = 1$, and from (\ref{e:prob-ratio}) we have
\begin{equation}
P_n(B_1, T_1) = P_n(B_2, T_2)
\end{equation}
in an adiabatic process.

\section{Quantum Otto engine cycle with strict adiabatic condition}

\begin{figure}[h]
\includegraphics[width=0.36\textwidth]{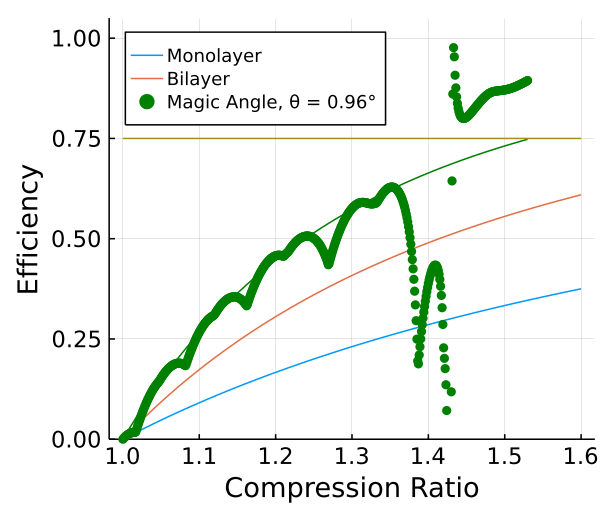}
\caption{\label{fig:otto2-efficiency} Efficiencies as a function of the
compression ratio for Otto cycles operating between $T_C = \SI{10}{K}$ and
$T_H = \SI{40}{K}$, and $B_1 = \SI{3.0}{T}$ with strict adiabatic
conditions. There is a divergence at compression ratio $=1.43$ as Work done changes sign, see Fig.~\ref{fig:otto2-work}. The green solid line is the least squares fit for case of MATBG. Obviously, efficiency is not defined for values of compression ratio beyond $1.43$ for MATBG.  }
\end{figure}
\begin{figure}[h]
\includegraphics[width=0.36\textwidth]{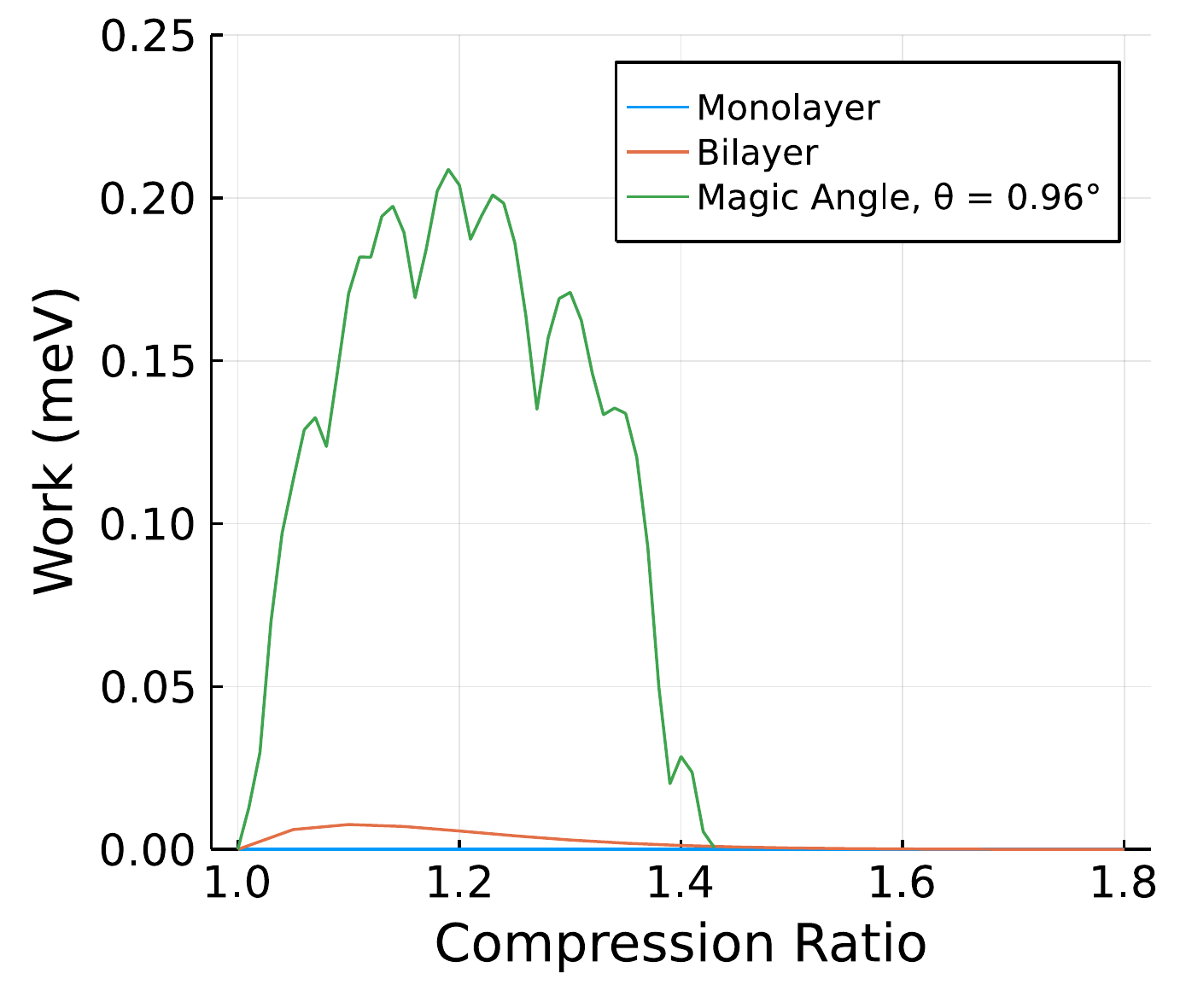}
\caption{\label{fig:otto2-work} Work outputs as a function of the
compression ratio for Otto cycles operating between $T_C = \SI{10}{K}$ and
$T_H = \SI{40}{K}$, and $B_1 = \SI{3.0}{T}$ with strict adiabatic
conditions.}
\end{figure}

\begin{table}[htp]
\caption{\label{table2}A comparison of efficiencies and work outputs for Otto
cycle operating with strict adiabatic conditions.}
\begin{ruledtabular}
\bgroup
\def\arraystretch{1.5}
\begin{tabular}{c|l d}
Working Substance & Efficiency & \multicolumn{1}{r}{Work (meV)} \\
\hline
Monolayer Graphene & $1 - r_C^{-1.00}$ & 0.000169\\
Bilayer Graphene & $1 - r_C^{-2.00}$ & 0.00758\\
Magic Angle Twisted Bilayer ($\theta^{\star}=0.96^{\circ}$) & $1 - r_C^{-3.24}$ & 0.204
\end{tabular}
\egroup
\end{ruledtabular}
\end{table}

Although the main text focuses on the heat engine cycle with general adiabatic strokes, a stricter version of the cycle can also be described very similarly herein. The two general adiabatic strokes are swapped out for stricter adiabatic strokes. We
require the stronger condition: $p_n(i) = p_n(f)$, and the cycle is identical
to Fig.~\ref{fig:otto-cycle} in all other respects.

We start with the graphene sample at temperature $T_C$ and Landau radius
$l_{B_1}$. Strict adiabatic compression implies, magnetic field is changed very
slowly $l_{B_1} \to l_{B_2}$, so that $\dd p_n = 0$ and the quantum condition
\begin{equation}
p_n(T_C, B_1) = p_n(2)
\label{e:qadiabatic-1}
\end{equation}
is satisfied. At the end of this stroke, the occupation probabilities will not
satisfy a Boltzmann distribution unless the energy levels change in the same
ratio, and the working substance will not, in general, be in a state with well
defined temperature.

The second stroke is the hot isochore; working substance is coupled with the heat
reservoir at temperature $T_H$, and the it absorbs heat. In
complete analogy to the classical case, heat exchanged is given by
\begin{equation}
Q_{2\to 3} = \sum_{n=0}^\infty E_n(B_2)[p_n(T_H, B_2) - p_n(2)].
\end{equation}

Next, in the adiabatic expansion $l_{B_2} \to l_{B_1}$, the magnetic field is
changed slowly so that the strict adiabatic condition
\begin{equation}
p_n(T_H, B_2) = p_n(4)
\label{e:qadiabatic-2}
\end{equation}
is satisfied. As before, the working substance does not have a well defined
temperature at the end of this stroke.
Finally, we have the cold isochore. The working substance is coupled with the
reservoir at temperature $T_H$ and the system returns to its initial state. As
before, the heat exchanged is given by
\begin{equation}
Q_{4\to 1} = \sum_{n = 0}^\infty E_n(B_1)[p_n(T_C, B_1) - p_n(4)],
\end{equation}
and we use the quantum first law of thermodynamics (\ref{e:quantum-flot}) to write
the work output
\begin{equation}
\vqty{W_\text{O}} = \vqty{Q_{2\to 3}} - \vqty{Q_{4\to 1}},
\end{equation}
and efficiency
\begin{equation}
\eta_\text{O} = 1 - \vqty{\frac{\sum_n E_n(B_1)[p_n(T_C, B_1) - p_n(4)]}
{\sum_n E_n(B_2)[p_n(T_H, B_2) - p_n(2)]}}.
\end{equation}
We use the above relations to compute efficiencies
(Fig.~\ref{fig:otto2-efficiency}) and work output (Fig.~\ref{fig:otto2-work})
in the Otto cycle with strict adiabatic conditions. To achieve optimal performance in
the quantum case, we take $T_C = \SI{10}{K}$, $T_H = \SI{40}{K}$ and $B_1
= \SI{3.0}{T}$. As the nature of adiabatic strokes differs between general and
strict cycles, it is not surprising that optimal performance is achieved for
different parameter values in both cases.

As in the general case, we do a
curve fitting for $\alpha$ in $\eta = 1 - r_C^-\alpha$ and find that $\alpha =
3.24$ for the magic angle. Results are summarized in Table~\ref{table2}. Although as compared to the general case, the efficiency in case MATBG has decreased a bit but is still much more efficient than mono or bi-layer. Work done in a strict QOE cycle has increased much more than the general cycle. Work done for the strict QOE cycle is almost $30$ times that seen in bilayer case, while that in general QOE cycle is $7$ times that in bilayer case.

\end{document}

%% file: preamble.tex
\usepackage{amsmath,amssymb}
\usepackage{bm}
\usepackage{dcolumn}
\usepackage{graphicx}
\usepackage[colorlinks=true,allcolors=blue]{hyperref}
\usepackage{IEEEtrantools}
\usepackage{lipsum}
\usepackage{microtype}
\usepackage{multirow}
\usepackage{physics}
\usepackage{siunitx}

\usepackage[T1]{fontenc}
\usepackage{newtxtext,newtxmath}

\newcolumntype{d}{D{.}{.}{-1}}

\providecommand{\dg}{^\circ}
\providecommand{\half}{\frac{1}{2}}
\providecommand{\inv}{^{-1}}
\providecommand{\nn}{\nonumber}